\documentclass[aps,pre,reprint,superscriptaddress,showpacs]{revtex4-1}

\usepackage{graphicx}
\usepackage{dcolumn}
\usepackage{bm}

\begin{document}


\title{Phase transitions of the $q$-state Potts model on multiply-laced Sierpinski gaskets}


\author{Liang Tian}
\affiliation{Beijing Computational Science Research Center, 3 Heqing
Road, Haidian District, Beijing 100080, China} \affiliation{College
of Science, Nanjing University of Aeronautics and Astronautics,
Nanjing 210016, China}
\author{Hui Ma}
\affiliation{Beijing Computational Science Research Center, 3 Heqing
Road, Haidian District, Beijing 100080, China}
\author{Wenan Guo}
\affiliation{Department of Physics, Beijing Normal University,
Beijing 100875, China}
\author{Lei-Han Tang}
\affiliation{Beijing Computational Science Research Center, 3 Heqing
Road, Haidian District, Beijing 100080, China}
\affiliation{Department of Physics, Hong Kong Baptist University,
Kowloon Tong, Kowloon, Hong Kong}



\date{\today}

\begin{abstract}
We present an exact solution of the $q$-state Potts model on a class
of generalized Sierpinski fractal lattices. The model is shown to
possess an ordered phase at low temperatures and a continuous
transition to the high temperature disordered phase at any $q\geq
1$. Multicriticality is observed in the presence of a
symmetry-breaking field. Exact renormalization group analysis yields
the phase diagram of the model and a complete set of critical
exponents at various transitions.

\end{abstract}


\maketitle

\section{Introduction}

The study of models on hierarchically organized lattices has
enriched our understanding of critical phenomena at phase
transitions~\cite{Berker,Gefen79,Griffiths81,Derrida83,DeSimoi}. In
many cases, the renormalization group (RG) transformation of the
partition function can be carried out
exactly~\cite{Dhar77,Berker,Griffiths84}. Exhaustive search of the
RG fixed points and a complete characterization of the flow in the
Hamiltonian space then become possible. Various conjectures on the
universality of the critical exponents and their dependence on
dimensionality and other model parameters can be tested. The
intuition gained from the study can be used to guide the
interpretation of numerical results and experiments.

In this work, we analyze an exact solution of the $q$-state Potts
model~\cite{Wu} on a class of fractal lattices with and without a
symmetry-breaking field. These lattices, first introduced by Menezes
and Magalh\~ases\cite{Menezes92}, are simple extensions of the
well-known Sierpinski gasket. They can be constructed iteratively
following the scheme shown in Fig. 1. The Potts model (including the
Ising model at $q=2$) on the original Sierpinski gasket does not
order except at zero
temperature~\cite{Gefen79,Gefen83,Luscombe,Borjan}. This may seem
surprising at first sight since the gasket has a fractal dimension
$d_{\rm f}=\ln 3/\ln 2=1.58$ greater than the lower critical
dimension $d_{\rm lc}=1$ from field theory. The puzzle was partially
resolved by Gefen, Mandelbrot and Aharony~\cite{Gefen79,Gefen84} who
noticed that, in addition to the fractal dimension, other
topological features of a lattice also affect the existence of a
finite temperature transition and its critical properties.

\begin{figure}
\scalebox{0.6}[0.6]{
\includegraphics{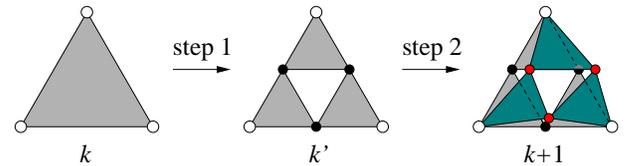}%
} \caption{\label{fig1}(Color online.) Iterative construction of the
multiply-laced Sierpinski gasket at $m=2$ in two steps and the
renormalization group transformation. Potts spins at the corners of
the triangle (open circles) are used to define the restricted
partition functions at a given order $k$. Solid circles indicate the
internal vertices to be integrated out to yield the renormalized
interactions.}
\end{figure}

The laced Sierpinski gasket can be considered as a hybrid of the
Sierpinski gasket and the Berker lattice, which does allow a finite
temperature transition with a Potts
Hamiltonian~\cite{Berker,Griffiths84}. As in the case of the Berker
lattice, the coordination number of corner sites on the lattice
grows with iteration. Presence of such vertices gives rise to a
hierarchical set of ``integration points'' that mediate order among
subsystems at low temperatures despite their tenuous contact. We
construct the phase diagram and calculate the critical properties
based on exact RG transformations derived from previous work by
Borjan et al.\cite{Borjan}. Interestingly, when a symmetry-breaking
field is introduced, we observe multicritical behavior and a line of
continuous transition between two disordered phases.

The paper is organized as follows. The fractal lattice and the Potts
model is introduced in Sec. 2. In Sec. 3 we present the iterative
relations for the partition function and discuss topological
features of the RG trajectories under the mapping, including the
relevant fixed points. Section 4 contains calculations of the phase
diagram and the critical exponents. A brief summary is presented in
Sec. 5.

\section{Potts model on laced Sierpinski gasket}

Figure 1 shows the iterative construction of the fractal lattice in
two steps. In the first step, the triangle at a given order $k$ is
triplicated and linked together at the three {\it $(k+1)$th order
internal vertices} (solid circles). In the second step, the
structure is replicated $m$-fold and then laced at the three {\it
$(k+1)$th order corner vertices} (open circles) to form a $(k+1)$th
order triangle. The zeroth order triangle is chosen to be a stack of
$m$ unit triangles laced at the three corner sites. Each pair of
sites are $m$-tuply connected.

Let $N_k, B_k$ and $T_k$ be the total number of sites, bonds and
basic triangles on a $k$th order triangle, respectively. From the
above definition, we have $N_{k+1}=3mN_k-6m+3, B_{k+1}=3mB_k$ and
$T_{k+1}=3mT_k$, with $N_0=3, B_0=3m$ and $T_0=m$. Hence
\begin{equation}
N_k={(3m)^{k+1}-1\over 3m-1}+2, B_k=(3m)^{k+1},T_k=m(3m)^k.
\label{total_number}
\end{equation}
Since the linear size of the triangle doubles upon each iteration,
the fractal dimension of the lattice is given by $d_{\rm
f}=\ln(3m)/\ln 2$.

The coordination number or the degree of a $k$th order corner vertex
is given by $C_k=2m^{k+1}$. The coordination number of an internal
vertex at the same order is $D_k=2C_{k-1}=4m^k$. On a fractal
lattice of order $n$, there are three $n$th order corner vertices.
The number of $k$th order internal vertices can be calculated from
the iterative relation $N_{n,k}=3mN_{n,k+1}$, with $N_{n,n}=3m$.
Hence $N_{n,k}=(3m)^{n-k+1}$. It is easily verified that
$N_n=3+\sum_{k=1}^nN_{n,k}$ and $B_n={3\over 2}C_n+{1\over
2}\sum_{k=1}^n N_{n,k}D_k$. Note that the number of internal
vertices with a given degree $D$ satisfies $N(D)\sim D^{-\kappa}$
where $\kappa =\ln(3m)/\ln(m)$. For the Sierpinski gasket at $m=1$,
all internal vertices have the same coordination number $D=4$.

We now assign a Potts variable $\sigma_i=0,1,\ldots,q-1$ to each of
the lattice sites. The energy of a Potts configuration
$\{\sigma_i\}$ is given by,
\begin{equation}
{\cal H}=-J\sum _{\langle ij\rangle} \delta(\sigma_i,\sigma_j)
-H\sum_i\delta(\sigma_i,0). \label{eq1}
\end{equation}
The first sum extends over all pairs of sites connected by a bond,
while the second sum
introduces a symmetry-breaking field that favors or disfavors a
selected Potts spin state $\sigma=0$. Here
$\delta(\sigma,\sigma')=1$ if $\sigma=\sigma'$ and 0 otherwise. In
this work we shall restrict ourselves to the ferromagnetic case at
$J>0$.

Model (\ref{eq1}) is not invariant under the RG transformation
discussed below. However, it can be recast in the form of an
invariant model that contains the one-spin term $H$ and a set of
three-spin (which includes the pair-wise) interactions on a
triangle. Making one spin state (e.g., $\sigma=0$) special, there
are altogether seven non-symmetry-related three-spin configurations
as shown in Fig. 2. At a given temperature $T$, the three-spin
interactions can be expressed in terms of Boltzmann weights
$\{Z_\alpha\}$ for the triplet configurations~\cite{Borjan}. The
one-spin term is represented by a weight $v=\exp(H/T)$ assigned to
all internal sites in the state $\sigma=0$. A {\it restricted
partition function} $Z_{\alpha,k}$ of the fractal lattice at order
$k$ is obtained by summing over spins on all internal sites, whereas
spins at the three corner sites define the label $\alpha$ according
to the scheme shown in Fig. 2.

\begin{figure}
\scalebox{0.5}[0.5]{
\includegraphics{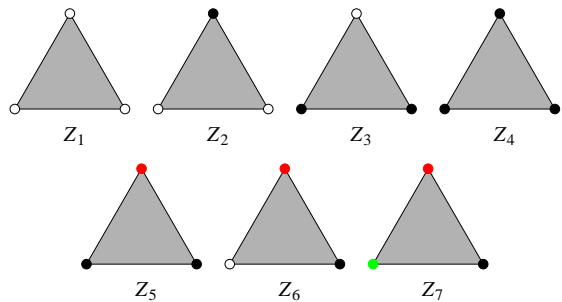}%
} \caption{\label{fig2}(Color online.) Seven non-symmetry-related
three-spin configurations and the associated Boltzmann weights (or
restricted partition functions). Open circles represent spins in the
selected state $\sigma=0$. Solid circles represent spins in the
remaining $q-1$ states distinguished by color.}
\end{figure}

\section{Iterative calculation of the partition function and the renormalization group flow}
The partition function of the Potts model with three-spin
interactions can be computed exactly by following the two-step
construction of Fig. 1. In the first step, $\{Z'_\alpha\}$ on the
lattice $k'$ are obtained from $\{Z_\alpha\}$ by summing over spins
on the three internal sites (solid circles). The mapping
$\{Z_\alpha\}\rightarrow \{Z'_\alpha\}$ for general $q$ was worked
out by Borjan et al., where $v$ enters as a parameter of the RG
transformation~\cite{Borjan,note1}. In the second step, each of the
seven weights are replicated as $Z'_\alpha\rightarrow \tilde
Z_\alpha=(Z'_\alpha)^m$.


To illustrate the procedure, we consider first the symmetric case
where the $\sigma=0$ state is made identical to the other $q-1$
states, i.e., $Z_{1,0}=Z_{4,0}$, $Z_{2,0}=Z_{3,0}=Z_{5,0}$,
$Z_{6,0}=Z_{7,0}$, and $v=1$. This includes model (\ref{eq1}) at
$H=0$. The RG transformation defines a two-dimensional map for the
variables $x\equiv Z_5/Z_4$ and $y\equiv Z_7/Z_4$. Let
$z_\alpha\equiv Z'_\alpha/Z_4^3$, the general iterative relations of
Ref. \cite{Borjan} reduces to:
\begin{eqnarray}
\label{z-iteration}
z_4(x,y)&=&1+(q-1)x^2(3+4x)\nonumber\\
&&+(q-1)(q-2)y\left[3x^2+3xy+(q-3)y^2\right],\nonumber\\
z_5(x,y)&=&x+4x^2+3x^3\nonumber\\
&&+(q-2)\left(2xy+y^2+5x^3+8x^2y+3x y^2\right)\nonumber\\
&&+(q-2)(q-3)y\left[x^2+7xy+(q-3)y^2\right],\\
z_7(x,y)&=&3x^2+6x y+8x^3+9x^2y+y^3\nonumber\\
&&+(q-3)\left(3y^2+x^3+21x^2y+9xy^2+3y^3\right)\nonumber\\
&&+(q-3)(q-4)y^2[9x+(q-2)y].\nonumber
\end{eqnarray}
The new ratios of the restricted partition functions after one
iteration are then given by,
\begin{eqnarray}
\tilde{x}&=&\phi(x,y)\equiv \Bigl[{z_5(x,y)\over z_4(x,y)}\Bigr]^m,\nonumber\\
&&\label{tilde-xy}\\
\tilde{y}&=&\psi(x,y)\equiv \Bigl[{z_7(x,y)\over
z_4(x,y)}\Bigr]^m.\nonumber
\end{eqnarray}

Specializing on the model (\ref{eq1}) at $H=0$, we have
$Z_{4,0}=\exp(3mJ/T), Z_{5,0}=\exp(mJ/T)$, and $Z_{7,0}=1$. The
starting values of the iteration are
$x_0=Z_{5,0}/Z_{4,0}=\exp(-2mJ/T)$ and
$y_0=Z_{7,0}/Z_{4,0}=\exp(-3mJ/T)$. Successive applications of Eqs.
(\ref{z-iteration}) and (\ref{tilde-xy}) yield the ratios
$x_{k+1}=\phi(x_k,y_k)$ and $y_{k+1}=\psi(x_k,y_k)$.

The full partition function on the $n$th order triangle is  obtained
from the restricted partition functions,
\begin{equation}\label{eq9}
 Z_k=q\left[1+3(q-1)x_k+(q-1)(q-2)y_k\right]Z_{4,k},
\end{equation}
where $Z_{4,k}$ satisfies the iterative relation,
\begin{equation}
Z_{4,k}=[z_4(x_{k-1},y_{k-1})Z_{4,k-1}^3]^m. \label{Z_k0}
\end{equation}
To compute the free energy $F_k=-T\ln Z_k$, we obtain from Eq.
(\ref{Z_k0}),
\begin{eqnarray}\label{eq12}
\ln  Z_{4,k}&=&m \ln  z_4(x_{k-1},y_{k-1})+3m \ln Z_{4,k-1}\nonumber\\
&=&B_kJ/T+\sum _{n=0}^{k-1} T_{k-n-1}\ln z_4(x_n,y_n).\label{eq12}
\end{eqnarray}
The free energy per unit triangle in the infinite size limit is thus
expressed as,
\begin{equation}\label{eq13}
f(T)\equiv \lim_{k\rightarrow\infty}{F_k\over T_k}=-3J-T\sum
_{n=0}^{\infty } (3m)^{-n-1}\ln  z_4(x_n,y_n).
\end{equation}


The mapping defined by Eqs. (\ref{z-iteration}) and (\ref{tilde-xy})
has two trivial fixed points at ${\rm L}=(0,0)$ and ${\rm H}=(1,1)$
on the $(x,y)$-plane, corresponding to the ordered and disordered
phases, respectively. As noted previously\cite{Gefen83}, on the
Sierpinski gasket at $m=1$, the mapping at small $x$ yields,
\begin{equation}\label{eq5}
    \tilde{x}\approx x+4x^2,
\end{equation}
which grows under the iteration. Therefore the low temperature fixed
point L is unstable and no finite temperature transition is
expected.

For $m>1$, we have $\phi(x,y)\simeq x^m$ at small $x$ which turns L
into a stable fixed point. Consequently, the Potts model on such a
lattice is ordered at sufficiently low
temperatures~\cite{Menezes92}. A third fixed point ${\rm
C}=(x_c,y_c)$ emerges and controls the critical properties at the
transition to the high temperature disordered phase. Figure 3 shows
the flow structure of the map at $m=2$ and $q=3$ which is
representative of the general situation. The thick solid line is the
critical manifold that separates the $(x,y)$-plane into two regions.
Below this line, the mapping carries the system to the low
temperature fixed point L. Hence this part of the parameter space is
associated with the ordered phase. Above the line, the mapping
carries the system to the high temperature fixed point H and the
system is disordered. On the critical manifold, the mapping carries
the system to the hyperbolic fixed point C.

\begin{figure}
\scalebox{0.4}[0.4]{
\includegraphics{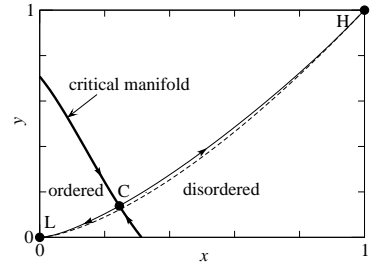}%
} \caption{\label{fig3}Fixed points (solid circles) and invariant
manifolds (solid lines) of the renormalization group transformation
at $m=2$ and $q=3$. The critical manifold (thick solid line) divides
the parameter space into high and low temperature regions. The
dashed line given by $y=x^{3/2}$ corresponds to the Potts model
(\ref{eq1}) at $H=0$.}
\end{figure}

In the neighborhood of the fixed point C, the mapping can be well
approximated by a linearized version applied to $\delta x_k\equiv
x_k-x_c$ and $\delta y_k\equiv y_k-y_c$,
\begin{equation}
\left(
\begin{array}{c}
\delta x_{k+1}\\
\delta y_{k+1}
\end{array}
\right)=\left(
\begin{array}{cc}
\partial_x\phi& \partial_y\phi\\
\partial_x\psi& \partial_y\psi
\end{array}
\right)\left(
\begin{array}{c}
 \delta x_k \\
 \delta y_k
\end{array}
\right), \label{eq21}
\end{equation}
where the derivatives are evaluated at $(x_c,y_c)$. Diagonalization
of the matrix yields two eigenvalues $\lambda_1=\lambda_T>1$ and
$\lambda_2<1$ that describe the expansion and contraction rates
along the unstable and stable manifolds of the hyperbolic fixed
point C, respectively.

More generally, using $Z_4$ as the normalization factor, the RG
transformation applied to the seven restricted partition functions
yields a nonlinear map in six dimensions spanned by $g_\alpha\equiv
Z_\alpha/Z_4$. At $v=1$, the symmetric case defines a
two-dimensional invariant manifold of the mapping, with L, H and C
the three fixed points on this manifold. By tracing the RG flow
along the unstable directions emanating from these points, we obtain
a total of 9 fixed points as listed in Table 1. Also given in the
table is the number of unstable directions $n_{\rm u}$ (known as the
codimension) and the maximum eigenvalue $\lambda_{\rm m}$ of the
linearized map at each fixed point.

Figure 4 illustrates schematically the RG flow between the fixed
points. Points labelled by L$'$, C$'$, H$'$ and L, C$''$, H$''$ lie
on the decoupled manifold with $g_2=g_3=g_6=0$. In this case,
$\tilde Z_1=Z_1^{3m}$ transforms by itself. The remaining three
weights $Z_4, Z_5$ and $Z_7$ transform according to Eqs.
(\ref{z-iteration}) and (\ref{tilde-xy}) with the replacement
$q\rightarrow q'=q-1$. The critical fixed point of the $(q-1)$-state
Potts model is denoted by $(x'_c,y'_c)$. At criticality, $Z'_4$
grows by a factor $(z'_{4,c})^m$ upon each iteration.

\begin{table}
\caption{\label{T1}Fixed points of the RG transformation at $v=1$.}
\begin{ruledtabular}
\begin{tabular}{c|cccccc|cc}
 &$g_1$&$g_2$&$g_3$&$g_5$&$g_6$&$g_7$&$n_{\rm u}$&$\lambda_{\rm m}$\\\hline
C&1&$x_c$&$x_c$&$x_c$&$y_c$&$y_c$&3&$\lambda_H$\\
H&1&1&1&1&1&1&1&$m$\\
\hline
C$''$&$(z'_{4,c})^{m\over 3m-1}$&0&0&$x'_c$&0&$y'_c$&2&$3m$\\
H$''$&$(q-1)^{3m\over 3m-1}$&0&0&1&0&1&1&$3m$\\
L&1&0&0&0&0&0&1&$3m$\\
C$'$&0&0&0&$x'_c$&0&$y'_c$&1&$\lambda'_T$\\
\hline
L$'$&0&0&0&0&0&0&0\\
H$'$&0&0&0&1&0&1&0\\
\hline
G&$\infty$&$\infty$&$\infty$&0&$\infty$&0&0\\
\end{tabular}
\end{ruledtabular}
\end{table}

\begin{figure}[b]
\scalebox{0.6}[0.6]{
\includegraphics{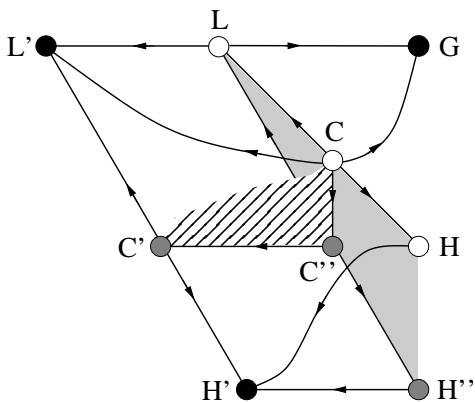}%
} \caption{\label{fig4}Topological organization of selected fixed
points and invariant manifolds under the renormalization group
transformation at $v=1$. Solid circles in black are stable fixed
points associated with the three bulk phases, separated by the
shaded and cross-hatched surfaces that represent first order and
continuous transitions, respectively. Open circles lie on the
invariant manifold where all $q$ spin states are identical. }
\end{figure}

The above RG flow topology is largely preserved in a non-zero field
$H$. The decoupled manifold at $g_2=g_3=g_6=0$ is invariant under
the RG transformation at any $v$. Among the two fixed points not on
this manifold, H is always a fixed point of the transformation. The
point C survives when $H\neq 0$, but its location shifts with the
broken symmetry.

The structure of the RG flow as shown in Fig. 4 is reminiscent of
the two-dimensional site-diluted three-state Potts model for the
physisorption of Kr on graphite studied by Berker et
al.~\cite{Berker78} The three globally stable fixed points L$'$,
H$'$ and G each represent a ``bulk phase'': L$'$ for the
$(q-1)$-state ordered or in-registry solid phase, H$'$ for the
$(q-1)$-state disordered or liquid phase, and G for the vacancy
($\sigma=0$) dominated gas phase. The shaded surface in Fig. 4
indicates the phase boundary between the condensed (L$'$ and H$'$)
and gaseous (G) phases. This transition is first order except on the
line CH, to be discussed below. On the other hand, transition
between the ordered and disordered condensed phases (cross-hatched
surface in Fig. 4), controlled by the $(q-1)$-state hyperbolic fixed
point C$'$, is continuous.

\section{Phase diagram of the Potts model and critical exponents}

The phase diagram of the Potts model defined by (\ref{eq1}) follows
from the general topological structure of the renormalization group
flow presented in the previous section. As usual, the nature of
ordering in the system at a given $T$ and $H$ is determined by which
of the three fixed points L$'$, H$'$ and G the RG transformation
carries the set of reduced partition functions to. From symmetry
considerations, the phase associated with G occupies the quadrant at
$H>0$. On the $H<0$ side, an order-disorder transition among the
$q-1$ favored Potts states takes place at a temperature $T_c(H)$.
The transition temperature at a given $H$ can be determined
numerically by noting that the system flows to L$'$ for $T<T_c(H)$
but to H$'$ for $T>T_c(H)$ under successive RG transformations. Note
that this transition is absent in the special case $q=2$.

Figure 5 shows the phase diagram of the Potts model for the
representative case $q=3$ and $m=2$. In the absence of the
symmetry-breaking field (i.e., $H=0$), spontaneous ordering occurs
for $T<T_{\rm bc}\simeq 2.8919$ with $q$-fold degeneracy. A positive
(negative) $H$ selects (de-selects) the $\sigma=0$ state. Thus, as
in the case of the Ising model at $q=2$, the line at $H=0$,
$T<T_{\rm bc}$ (solid line in Fig. 5) becomes a first order
transition line on the $T$-$H$ plane. The order-disorder transition
continues on the $H<0$ side (dashed line in Fig. 5) and joins the
line of first order transition tangentially at the bicritical point
$(T_{\rm bc}, H=0)$\cite{Fisher_Nelson}. Furthermore, the gas and
the $(q-1)$-state disordered phases are separated by a continuous
transition on the high temperature side.

\begin{figure}
\scalebox{0.4}[0.4]{
\includegraphics{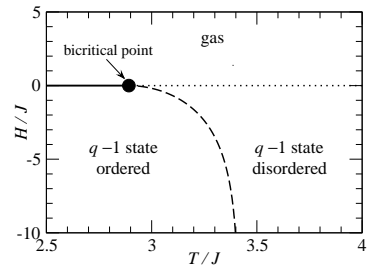}%
} \caption{\label{fig5}Phase diagram of the Potts model (\ref{eq1})
at $m=2$ and $q=3$. Lines of first order (solid) and continuous
(dashed and dotted) transitions meet at the bicritical point $H=0$,
$T=T_{\rm bc}$.}
\end{figure}

We now turn to the calculation of the critical exponents that
describe the singular behavior of various thermodynamic quantities
at the continuous phase transitions. The standard RG procedure
allows them to be computed from the eigenvalues of the linearized
map at the appropriate fixed points. For example, the order-disorder
transition at $H<0$ (dashed line in Fig. 5) is governed by the fixed
point C$'$ which has a single unstable direction with an eigenvalue
$\lambda'_T(q)\equiv\lambda_T(q-1)$. In the neighborhood of the
transition point $T=T_c(H)$, the free energy contains a singular
part $f_s\sim |T-T_c(H)|^{2-\alpha'}$. The standard scaling analysis
yields the specific heat exponent $\alpha'=2-\ln(3m)/\ln\lambda'_T$
for the $(q-1)$-state Potts model, independent of $H$.

The first order transition at $H=0, T<T_{\rm bc}$ is governed by the
fixed point L. The usual scaling argument suggests that the singular
part of the free energy behaves as $f_s\sim |H|^{1+1/\delta}$, where
$1+1/\delta = \ln d_f/\ln \lambda$, with $\lambda$ being the
eigenvalue for the unstable direction of the RG flow. At L,
$\lambda=3m=d_f$ gives $\delta=\delta_{\rm L}=\infty$, hence
\begin{equation}
f_s\sim |H|,\qquad (T<T_{\rm bc}). \label{T_small}
\end{equation}
Equation (\ref{T_small}) can also be derived from the jump in the
fraction $p_0$ of vertices in the $\sigma=0$ state across the
transition line. At $H=0$, we have ``phase-coexistence'' of the $q$
spontaneous symmetry-breaking states, each favoring a particular
Potts state whose fraction on the lattice $p(T)>1/q$. A positive
field $H$ selects the $\sigma=0$ state with $p_0=p(T)$. On the other
hand, a negative $H$ de-selects the $\sigma=0$ state, yielding
$p_0=[1-p(T)]/(q-1)$. In each case, the free energy changes from its
$H=0$ value by an amount $\Delta f =-Hp_0$ per vertex. The jump in
$p_0$ gives rise to the singularity in the free energy above.

For $T>T_{\rm bc}$, dependence of the free energy on $H$ is
controlled by the RG flow near the fixed point H. Table 1 gives
$\lambda=m$ and hence $\delta=\delta_{\rm
H}=1/[\ln(3m)/\ln(m)-1]=\ln m/\ln 3$. Consequently,
\begin{equation}
f_s\sim |H|^{1+(\ln 3/\ln m)},\qquad (T>T_{\rm bc}). \label{T_large}
\end{equation}
Such a singular behavior on the high temperature side is quite
unusual. The common notion is that the disordered system consists of
statistically independent subsystems of finite size. Since the free
energy of each subsystem is analytic in the controlling parameters
$T$ and $H$, the free energy of the system as a whole should be
analytic as well. However, examining Eq. (\ref{eq1}) more closely,
we find an interesting possibility. Consider a $k$th order vertex
and its $C_k=2m^{k+1}$ neighbors. A positive external field $H>0$
generates an excess probability of order $H$ for each neighboring
spin to be in the $\sigma=0$ state. This gives rise to an effective
field $H_{{\rm eff},k}\sim HC_k\simeq Hm^k$ at the vertex in
question from the ferromagnetic couplings, which grows exponentially
with $k$. Setting $H_{\rm eff}=T$ yields a critical value $k_c=
-\ln(H/T)/\ln m$. Vertices with $k>k_c$ are completely ordered.
Contribution to the free energy per triangle from these vertices is
estimated to be $f_s\sim T_{k_c}^{-1}H_{{\rm eff},k_c}\sim
(3m)^{-k_c}\sim H^{\ln(3m)/\ln m}$ which agrees with Eq.
(\ref{T_large}).

We now consider the most complex situation near the bicritical point
at $H=0, T=T_{\rm bc}$. The RG flow is controlled by the fixed point
C which has three unstable directions as shown in Fig. 4. In terms
of the scaling fields $t=(T-T_{\rm bc})/T_{\rm bc}$ along CH, $H$
along CG, and $H_C$ along CC$''$, the singular part of the free
energy is expected to satisfy the scaling,
\begin{equation}
f_{\rm s}(t,H, H_C)=b^{-d_{\rm f}}f_{\rm
s}(b^{y_T}t,b^{y_H}H,b^{y_C}H_C), \label{homo_f}
\end{equation}
where $b$ is an arbitrary scaling factor. The scaling exponents
$y_a$ are related to the eigenvalues $\lambda_a$ of the linearized
map through $y_a=\ln\lambda_a/\ln 2$. For the model defined by Eq.
(\ref{eq1}), $H_C$ can be expressed as a linear combination of $t$
and $H$ but since $H_C=0$ when $H=0$, we may write $H_C=H$. With the
choice $b=|t|^{-1/y_T}$, Eq. (\ref{homo_f}) takes a more suggestive
form,
\begin{equation}
f_{\rm s}(t,H)=|t|^{2-\alpha}{\cal F}_\pm\Bigl({H\over
|t|^\Delta},{H\over |t|^\phi}\Bigr), \label{f_scaling}
\end{equation}
where the subscript '$\pm$' indicates the sign of $t$. Here
$\alpha=2-\ln(3m)/\ln(\lambda_T)$,
$\Delta=\ln\lambda_H/\ln\lambda_T$, and
$\phi=\ln\lambda_C/\ln\lambda_T$ are the specific heat, gap, and
crossover exponents, respectively.

Figure 6 shows the numerically computed eigenvalues $\lambda_H,
\lambda_T$, $\lambda_C$ and the exponents $\alpha, \Delta, \phi$ at
$m=2$ but different $q$ values. It is evident that
$\lambda_H>\lambda_T>\lambda_C$ and $\Delta>1>\phi$ for all $q$.
Dominance of the symmetry-breaking field renders most of the
neighborhood of the bicritical point to be occupied by either the
$(q-1)$-state ordered phase at $H<0$ or the gas phase at $H>0$. The
$(q-1)$-state disordered phase appears in a narrow power-law wedge
on the $H<0$ side, bounded below by the transition line
 $H=H_c(T)\simeq -At^\Delta$. Within this phase, $H/t^\phi\leq At^{\Delta-\phi}\rightarrow 0$
 as $t\rightarrow 0$. The scaling function ${\cal F}_+(u,w)$ in Eq. (\ref{f_scaling})
is singular on the two transition lines at $u=0$ and $u=-A$,
respectively. In the latter case, the scaled variable $w=H/t^\phi$
controls the crossover from the $q$-state to the $(q-1)$-state Potts
criticality. Details will not be elaborated here.

\begin{figure}
\scalebox{0.4}[0.4]{
\includegraphics{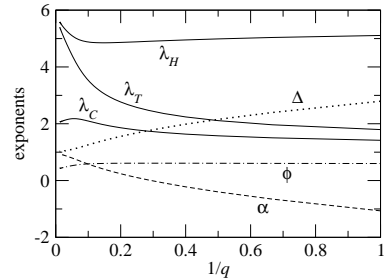}%
} \caption{\label{fig6}Eigenvalues $\lambda_H, \lambda_T$ and
$\lambda_C$ associated with the three unstable directions of the
renormalization group flow at C against $1/q$. Here $m=2$. Also
shown are the specific heat exponent $\alpha$ (dashed line), the gap
exponent $\Delta$ (dotted line), and the crossover exponent $\phi$
(dash-dotted line).}
\end{figure}

At large $q$ values, one may approximate the mapping near C with
$\phi(x,y)\simeq\psi(x,y)\simeq (qy)^{3m}$ which yields $x_c\simeq
y_c\simeq q^{-3m/(3m-1)}$. Consequently,
$\lambda_T(q\rightarrow\infty)=3m$, in agreement with the trend seen
in Fig. 6. In this limit, C eventually merges with L, yielding
$\lambda_H=3m$ as well. Both $\alpha$ and $\Delta$ approach 1. The
magnetization exponent $\beta=2-\alpha-\Delta\rightarrow 0$. We have
checked that the above behavior is quite general for any $m>1$.

\section{Summary}

In this paper, we analyzed in detail solution of the Potts model on
a fractal lattice that has a low temperature ordered phase. Compared
to the Sierpinski gasket at $m=1$, the coordination number of
vertices at the corners of the fractal triangle with $m>1$ grows
with lattice size. Thus our findings are consistent with previous
understanding that a finite $T_c$ requires sufficiently strong
connectivity on the fractal lattice.

Exact renormalization group transformation of the restricted
partition functions is presented here with the help of previous work
by Borjan et al.\cite{Borjan} Although the mapping in the
six-dimensional space contains many fixed points and invariant
manifolds, a relatively simple RG flow diagram is obtained. The flow
is mainly defined by three stable fixed points, each representing a
bulk phase with a distinct symmetry or broken symmetry. The topology
of the RG flow is robust against the lattice geometry parameter $m$
and the number of Potts states $q$. By combining analytical and
numerical approaches, we have computed the phase diagram of the
model and a complete set of critical exponents for the transitions,
including those at the bicritical point, based on the standard RG
framework.

It is worth noting that, on the laced Sierpinski gasket, transition
to the high temperature disordered phase is continuous for all $q$
values. Previously, solution of the Potts model on the Berker
lattice also yielded continuous transitions with a specific heat
exponent $\alpha(q)$ that approaches 1 in the limit
$q\rightarrow\infty$~\cite{Berker,Griffiths84}. This is in contrast
to the $q$-state Potts model on periodic lattices in two and higher
dimensions, where the transition becomes first order when $q$
exceeds a critical value $q_c$. Further work is needed to gain a
precise understanding of this difference.

\begin{acknowledgments}
The work is supported in part by the Research Grants Council of the
Hong Kong SAR under grant 202309. L. Tian thanks the Hong Kong
Baptist University for hospitality where part of the work was
carried out.
\end{acknowledgments}

\nocite{*}


\begin{thebibliography}{}
\bibitem{note1} The variable $t$ in Eqs. (2)-(8) of Ref. \cite{Borjan}
is replaced by the vertex weight $v=\exp(H/T)=1/t$ for the state
$\sigma=0$. Note also the following typos in their expressions: i)
Eq. (2), coefficient of the last term should read $q^3-6q^2+11q-6$.
ii) Eq. (5), first term on the right-hand-side should read $Z_4^3$.
\end{thebibliography}

\end{document}